\documentclass[11pt,a4paper]{article}

\usepackage{epsfig}
\usepackage{amsmath}
\usepackage{graphicx}
\usepackage{float}
\usepackage{subfig}
\usepackage{vmargin}
\usepackage[noblocks]{authblk}

\usepackage[T1]{fontenc}

\newcommand{\bt}{\ensuremath{\beta}}

\newcommand{\be}{\begin{equation}}
\newcommand{\ee}{\end{equation}}
\newcommand{\ba}{\begin{equation} \begin{aligned}}
\newcommand{\ea}{\end{aligned} \end{equation}}

\newcommand{\SIR}{{\it SIR}}
\newcommand{\SEIR}{{\it SEIR}}

\title{Epidemic prediction and control in clustered populations}
\author[1,*]{Thomas House}
\author[1,2]{Matt J Keeling}
\affil[1]{Warwick Mathematics Institute, University of Warwick, Gibbet Hill
Road, Coventry, CV4 7AL, UK.}
\affil[2]{School of Life Sciences, University of Warwick, Gibbet Hill
Road, Coventry, CV4 7AL, UK.}
\affil[*]{Corresponding Author: T.A.House@warwick.ac.uk}

\date{}

\begin{document}

\maketitle

\begin{abstract}

There has been much recent interest in modelling epidemics on networks,
particularly in the presence of substantial clustering. Here, we develop
pairwise methods to answer questions that are often addressed using epidemic
models, in particular: on the basis of potential observations early in an
outbreak, what can be predicted about the epidemic outcomes and the levels of
intervention necessary to control the epidemic? We find that while some results
are independent of the level of clustering (early growth predicts the level of
`leaky' vaccine needed for control and peak time, while the basic reproductive
ratio predicts the random vaccination threshold) the relationship between other
quantities is very sensitive to clustering.

\end{abstract}

\section*{Introduction}

There has been much recent interest in modelling infectious diseases on contact
networks~\cite{Keeling:2005,Bansal:2007}. The incorporation of population
structure that deviates from homogeneous mixing is most generally
conceptualised as a network of epidemiologically relevant contexts, and an
increasing amount of data is available on these contacts, either from
surveys~\cite{Mossong:2008} or socio-demographic data~\cite{Eubank:2004}.  One
observed feature of realistic contact networks is the presence of an
appreciable number of short closed loops in the network, often called
\textit{clustering}.

Through construction of networks with special structure, it has recently been
shown possible to derive some exact results for clustered networks based on
households~\cite{Ball:2009,Ball:2010} and non-overlapping
triangles~\cite{Newman:2009,Miller:2009}. It has also been possible to write
down and integrate dynamical systems that capture transient epidemic behaviour
on such networks~\cite{Volz:2010,House:2010}.

At the same time, a more established approach to epidemics on clustered
networks exists in the form of pairwise equations, where an approximation is
made to produce a system of ordinary differential equations (ODEs) that depend
on a small number of real-valued parameters for the
network~\cite{Keeling:1999}. These approximations have been consistently found
to be in good qualitative agreement with stochastic
simulations~\cite{House:2010a}, and the lack of rigorous justification for the
closure used is compensated for in several ways.  For example, a system of ODEs
can easily be integrated, involves few parameters, and has a rigorous
definition of critical intervention thresholds.  There is also the benefit that
analytic manipulations can be performed on the equations involved to aid
theoretical understanding~\cite{Trapman:2007}.

One of the key uses of epidemic modelling is to predict outcomes and critical
intervention thresholds on the basis of observables, which can be done
analytically for the special case of household models~\cite{Goldstein:2009}.
In this paper we consider the insights that can be gained for epidemic dynamics
on networks with a general clustered structure from consideration of pairwise
equations. We start by reviewing standard epidemic and network theory. We then
outline our pairwise methodology, including the definition of observable
reproduction numbers and control thresholds. We then consider the implication
of this for intervention and outcome prediction.

\subsection*{The SIR model}

One of the simplest approaches to the problem of epidemic prediction is the
\SIR{} model, developed over 80 years ago~\cite{Kermack:1927}. This basic
paradigm has been extended and applied fruitfully to a wide range of
diseases~\cite{AndersonMay1991}.  In the absence of births and deaths, the
dynamics for this model can be written in terms of the proportions of the
population susceptible, $S(t)$, infectious, $I(t)$, and removed, $R(t)$, as
\ba
 \dot{S}(t) & = - \bt S(t) I(t) \text{ ,} \\
 \dot{I}(t) & = \bt S(t) I(t) - \gamma I(t) \text{ ,} \\
 \dot{R}(t) & = \gamma I(t) \text{ .}
\label{eq:sir}
\ea
This model has two parameters: $\gamma$, the rate of recovery from the
infectious state; and $\bt$, the rate at which new cases are created when
susceptible and infectious individuals interact.  An important quantity that
emerges from analysis of many epidemic models is the \emph{basic reproductive
ratio}, $R_0$, which we define (standardly) as the average number of secondary
cases produced by an average infectious individual early on, once the epidemic
dynamical system has settled onto its dominant eigenvector.  In the simple
\SIR{} model given by equations~\ref{eq:sir}, this quantity as defined is equal
to $\bt / \gamma$. 

Strictly speaking, since the system is non-linear, we should also consider
$I(0)$, the initial proportion of the population that is infectious, and
$S(0)$, the initial proportion of the population that is susceptible (since
they are proportions, the dynamical variables obey $S+I+R=1$ at all times). In
this work, we take $I(0) \ll 1$, and furthermore assume $S(0) \approx 1$.
Given these assumptions, for \SIR{} dynamics in a homogeneous population with
arbitrary recovery time distribution, the final size of the epidemic,
$R_{\infty}$, will be given by a solution to the transcendental equation
below~\cite{Kermack:1927}:
\be
 R_{\infty} = 1 - e^{- R_0 R_{\infty}} \text{ .}
\label{eq:rinf}
\ee
We can also write down exact expressions for the peak height and time in the
simple \SIR{} model, which are
\be
 I_* = 1 - \frac{1}{R_0} \left( 1 + \ln (R_0) \right) \text{ ,}\qquad
 t_* = \int_{0}^{1- \frac{1}{R_0} -I_*} \frac{dR}{1 - R - e^{-R_0 R} } 
  \quad \text{.}
\ee
It is another relevant standard result that, if a small amount of infection
$I(0)$ is introduced into an otherwise susceptible population then this model
predicts that at early times, the proportion of infectious individuals in the
population will be given by $I(t) \approx I(0) e^{\gamma(R_0
-1)t}$~\cite{AndersonMay1991}.

\subsection*{Network theory}

We consider a network of $N$ nodes labelled $i,j\ldots$ to be defined by its
adjacency matrix $G_{ij}$, which takes the value 1 when nodes $i,j$ are
connected and the value 0 otherwise. We define this also to be symmetric and
without self-edges, so $G_{ij} = G_{ji}$ and $G_{ii}=0$.

Since we are considering the impact of clustering, we try to remove the
impact of node degree heterogeneity, which can confound the effects of
clustering~\cite{Kiss:2008}, by assuming that every individual has the
same neighbourhood size $n$
\be
\forall i, \sum_j G_{ij} = n \text{ .}
\ee
It is worth noting that there is no reason why our methodology should not be
extended to heterogeneous degree distributions; our choice of neighbourhood
regularity is simply to consider the impact of clustering independently of
other network statistics.  The clustering coefficient $\phi$ is defined by
\be
\phi = \frac{\sum_{i,j,k} G_{ij} G_{jk} G_{ki}}%
{\sum_{i,j,k} G_{ij} G_{jk} (1 - \delta_{ik})} \in [0,1] \text{ ,}
\ee
where $\delta_{ik}$ is the Kronecker delta.  If the dynamical state of a node
$i$ is indicated by $A_i$, then we also use a notation where:
\be
[A] = \sum_i A_i \text{ ,}\quad
[AB] = \sum_{i,j} A_i B_j G_{ij} \text{ ,}\quad
[ABC] = \sum_{i,j,k} A_i B_j C_k G_{ij} G_{jk} \text{ .}
\ee

\section*{Methods}

\subsection*{Pairwise equations}

The pairwise equations for an \SIR{} epidemic are standardly given by
\ba
\dot{[S]} & = - \tau [SI] \text{ ,} \\
\dot{[I]} & = \tau [SI] - \gamma[I] \text{ ,} \\
\dot{[R]} & = \gamma[I] \text{ ,} \\
\dot{[SS]} & = - 2 \tau [SSI] \text{ ,} \\
\dot{[SI]} & = \tau ( [SSI] - [ISI] - [SI] ) -\gamma [SI] \text{ ,} \\
\dot{[II]} & = 2 \tau ( [ISI] + [SI]) -2 \gamma [II] \text{ ,} \\
\dot{[SR]} & = -\tau [ISR] + \gamma [SI] \text{ ,} \\
\dot{[IR]} & = \tau [ISR] + \gamma ( [II] - [IR]) \text{ ,} \\
\dot{[RR]} & = \gamma [IR] \text{ .} \\
\label{pw}
\ea
These equations are exact but unclosed.  While we restrict ourselves to the
simple \SIR{} model, it is possible to include more complex disease natural
histories such as the \SEIR{} model presented in~\cite{Rand:1999}.  To close
this system, we approximate the triples using the standard approximation for
clustered systems~\cite{Keeling:1999},
\be
 [ABC] \approx \frac{n-1}{n} \frac{[AB][BC]}{[B]} \left( (1-\phi) 
  + \phi \frac{N}{n} \frac{[CA]}{[A][C]} \right) \text{ .}
\label{close}
\ee
While there are several ways to recover the standard \SIR{} model from network
systems, for the pairwise approach the most natural limit is to take
$n\rightarrow\infty$ while holding $\beta = n \tau$ constant. We therefore
compare the scaled transmission rate $(n\tau / \gamma)$ to other quantities,
since this converges to all other reproduction numbers in the appropriate
limit. Another quantity that converges to other reproduction numbers in the
homogeneous limit is $n\tau/(\tau + \gamma )$, which is $R_0$ for an
unclustered network. Since this is a closed function of the scaled transmission
rate, we do not present additional results, but note that it is a quantity that
may well be estimated or inferred from data.  The system of
equations~\eqref{pw} can be numerically integrated using standard methods such
as Runge-Kutta, although our experience is that sophisticated solvers capable
of switching to stiff methods are significantly more accurate and numerically
efficient.

The closure~\eqref{close} can be used for any graph where the parameters $N$,
$n$ and $\phi$ are known. This closure is intended to be applied to graphs
where each node has degree close to $n$, although closures appropriate to more
heterogeneous systems do exist~\cite{House:2010a}.  Since the first application
of this closure to epidemic systems, attempts have been made to provide
rigorous justification of their validity~\cite{Rand:1999}. While there are no
exact results at present, from previous work we expect that the agreement is
likely to be best for regular (degree-homogeneous) configuration-model networks
where a small amount of clustering has been introduced through
rewiring~\cite{House:2010a}, and that agreement will be poor in the presence of
degree heterogeneity~\cite{Eames:2002,Bansal:2007}, where shortest path lengths
are long~\cite{Sharkey:2008} or when the network motif structure is not well
captured by the single parameter $\phi$~\cite{House:2009}.

\subsection*{Solution by linearising Ansatz}

One way to gain analytic traction on equations~\eqref{pw} is through
linearisation of the system, representing the situation early in the epidemic
when the number of susceptibles has not been significantly depleted. A
straightforward method is to define an Ansatz representing the intuition that
all dynamical variables should have their behaviour determined by the
prevalence of infection, and then to confirm that this Ansatz is indeed a
consistent solution.

We start by defining the early growth reproduction number $r_0$ through early
asymptotic growth in the proportion of infectious individuals. 
\be
I(t) =: I_{\rm start} e^{\gamma(r_0 -1)t} \text{ ,}
\ee
where $ I_{\rm start}$ is the proportion of infectious individuals at the start
of the period of exponential growth.  $r_0$ can therefore be measured by
fitting an exponential curve to the early growth in the number of cases at the
start of an epidemic, e.g.~\cite{Chowell:2007}.  Where $ I_{\rm start} \ll 1$,
we propose linearisation of the system of epidemic equations using the
following Ansatz:
\ba
 \dot{I}(t) & = \gamma (r_0 -1) I(t) \text{ ,}\\
 [A] & = [A]_0 + k_A I(t) \text{ ,}\\
 [AB] & = [AB]_0 + k_{AB} I(t) \text{ .}
 \label{linans}
\ea
Putting these substitutions into the system~\eqref{pw} closed by~\eqref{close}
and solving algebraically for $\{ r_0 , k_{A}, k_{AB} \}$ at given $\tau,
\gamma,n$ and $\phi$ allows us to parameterise the dynamical system.
Algorithmically, this offers significant advantages over, say, an iterative
scheme that involves repeated integration and modification of underlying
parameters to match an early growth curve.

\subsection*{Definition of $R_0$}

The primary difficulty of defining a `true' basic reproductive ratio that is
both a threshold and corresponds to the verbal definition in a clustered
population is that, even early in the epidemic, infectious individuals share
susceptible contacts and this competing risk does not allow the argumentation
used for locally tree-like networks or well-mixed populations.

\begin{figure}[H]
\begin{center}
\scalebox{0.95}{\resizebox{\textwidth}{!}{ \includegraphics{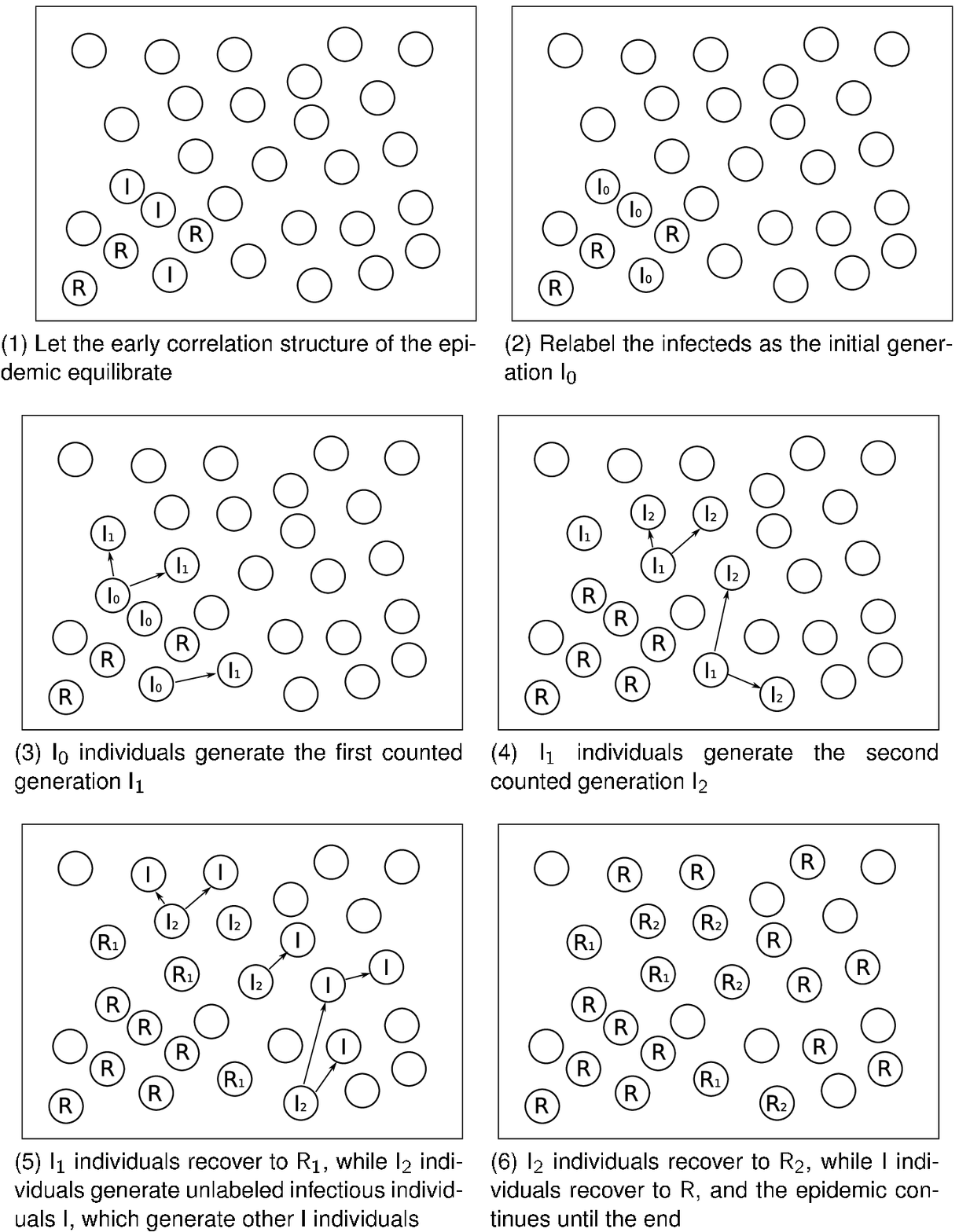} }}
\end{center}
\caption{Calculating the actual basic reproduction number through generation
counting. Susceptible individuals are not labelled for clarity.  The basic
reproductive number $R_0$ is given by the value of $R_2 / R_1$ at the end of
the epidemic.}
\label{fig:exp}
\end{figure}

In~\cite{Trapman:2007}, reproduction numbers were defined analytically for
clustered systems based on moment closure. We consider a different,
complementary method that can be applied to any epidemic model (including
stochastic and individual-based simulations) based on generation counting, but
which is numerical rather than analytic. Figure~\ref{fig:exp} shows the method
schematically. Firstly, the system is run until the early network correlation
structure has equilibrated and the dynamical system sits on its dominant
eigenvalue. We then relabel all infecteds as 0-th generation $I_0$, and label
two subsequent generations of infection $I_1,I_2$, which recover to $R_1,R_2$
respectively, before letting the epidemic proceed until no infectious
individuals of type $I_1$ and $I_2$ remain. The basic reproduction number $R_0$
is given by the final value of $R_2 / R_1$. To demonstrate this technique
applied to the standard \SIR{} model, consider the linearised equations
\be
\dot{I}_0 = -\gamma I_0 \text{ ,} \quad
\dot{I}_1 = \beta I_0 -\gamma I_1 \text{ ,} \quad
\dot{I}_2 = \beta I_1 -\gamma I_2 \text{ .}
\ee
The solution to this system is
\be
{I}_0 = I(0) e^{-\gamma t} \text{ ,} \quad
{I}_1 = I(0) \beta t e^{-\gamma t} \text{ ,} \quad
{I}_2 = \tfrac{1}{2} I(0) (\beta t)^2  e^{-\gamma t} \text{ ,}
\ee
which reproduces the standard $R_0 = \beta / \gamma$ when the ratio of the area
under the $I_2$ and $I_1$ curves is evaluated in the limit
$t\rightarrow\infty$.

There are, of course, some epidemic models such as simulations based on
lattices or highly complex individual behaviour where there is not an obvious
dynamical system underlying the model, and early behaviour of the epidemic does
not involve exponential growth. In these systems, the method of generation
counting will still give an answer that closely matches the standard definition
of $R_0$~\cite{Diekmann:1990}, but with spatial structure playing a comparable
role to risk structure.  For example, if a disease is sufficiently
transmissible to invade a square lattice then the quantity $R_2/R_1$ will
asymptote to unity, as would be expected of $R_0$.  Furthermore, if there is a
phase in a system's early dynamics (before the proportion of susceptibles in
the whole population has been significantly reduced) where the ratio $R_2/R_1$
reaches quasi-equilibrium, then this constant quantity will provide a threshold
for epidemic invasion.

\subsection*{Vaccination}

The parameters $n\tau / \gamma$, $r_0$ and $R_0$ as defined above are
observable early in an epidemic, but do not directly lead to critical
intervention thresholds for network epidemic models in the same way as in the
standard \SIR{} model.

In this paper, we consider two distinct interventions: reducing transmission
by a proportion $\varepsilon$ so $\tau \rightarrow (1-\varepsilon)\tau$,
and placing a proportion of the population $v$, chosen randomly at the
individual level, in the recovered class at the start of the epidemic, so that
\be
[S]_0 = (1-v) \text{ ,} \qquad [SS]_0 = (1-v)^2 \text{ .}
\ee
We preserve the degree distribution of the network by placing nodes in the
dynamically inert recovered class, so that e.g.\ $[R]_0 = v$; this is in
contrast to modelling vaccination by modification of the network topology,
and allows~\eqref{close} to remain valid.
The critical values sufficient to contain an epidemic, $\varepsilon_c$ and
$v_c$ can be calculated by the use of linearising Ansatz as above, then finding
values at which the predicted $r_0(v,\varepsilon)$ is 1. So that we are
comparing similar quantities, we use these critical values to define `leaky'
and `vaccination' reproduction numbers
\be
R_L = \frac{1}{1-\varepsilon_c} \text{ ,} \qquad
R_V = \frac{1}{1-v_c} \text{ .}
\ee
This terminology is taken from household models, where analytic results can
be obtained relating different reproduction numbers~\cite{Goldstein:2009}.

\section*{Results}

\subsection*{Analytic results}

We start by considering some analytic results that can be obtained from the
pairwise system. Our methods are developed with numerical solution in mind,
however some closed expressions can be derived by substituting~\eqref{close}
and~\eqref{linans} into~\eqref{pw}.  At small clustering in a regular graph
with $n$ links per node we can calculate the first-order impact of clustering
on $r_0$,
\be
r_0 =
 \frac{\tau}{\gamma} \left(
 (n-2)
 -\frac{2 (n-1) \left( 2(n-1)(n-2) \tau + n \gamma \right)}%
  {n^2 ((n-2) \tau +\gamma )}\phi  +O\left(\phi ^2\right)
 \right) \text{ ,} \label{linphir0}
\ee
which demonstrates the standard result that, leaving other parameters constant,
clustering reduces epidemic potential. Unclosed expressions to all orders in
$\phi$ can be found in~\cite[Eqns.\ (14, 19) with $r_* \rightarrow
r_0$]{Trapman:2007}.  In structured populations, there is a conceptual
difference between an intervention that reduces transmission by a fraction
$\varepsilon$ and a random, completely protective, vaccination of a proportion
$v$ of the population.  In the absence of clustering, this is reflected in the
difference between leaky and vaccinated early growth reproduction numbers as
below
\be
r_0(\varepsilon) = (1-\varepsilon )\frac{\tau}{\gamma}(n-2) \text{ ,}\qquad
r_0(v) = \frac{\tau}{\gamma}((n-1)(1-v) -1) \text{ .}
\ee
Interestingly, where $\phi =0$, this means that $R_L = r_0$, and also
\be
R_V = \frac{1}{1-v} = (n-1)\frac{\tau}{\tau + \gamma} = R_0 \text{ .}
\ee
In the clustered case these relationships between observables and
intervention-related thresholds do not hold exactly, however we will show that
they can remain numerically close.

\subsection*{Reproduction numbers}

\begin{figure}[H]
\begin{center}
\scalebox{0.95}{\resizebox{\textwidth}{!}{ \includegraphics{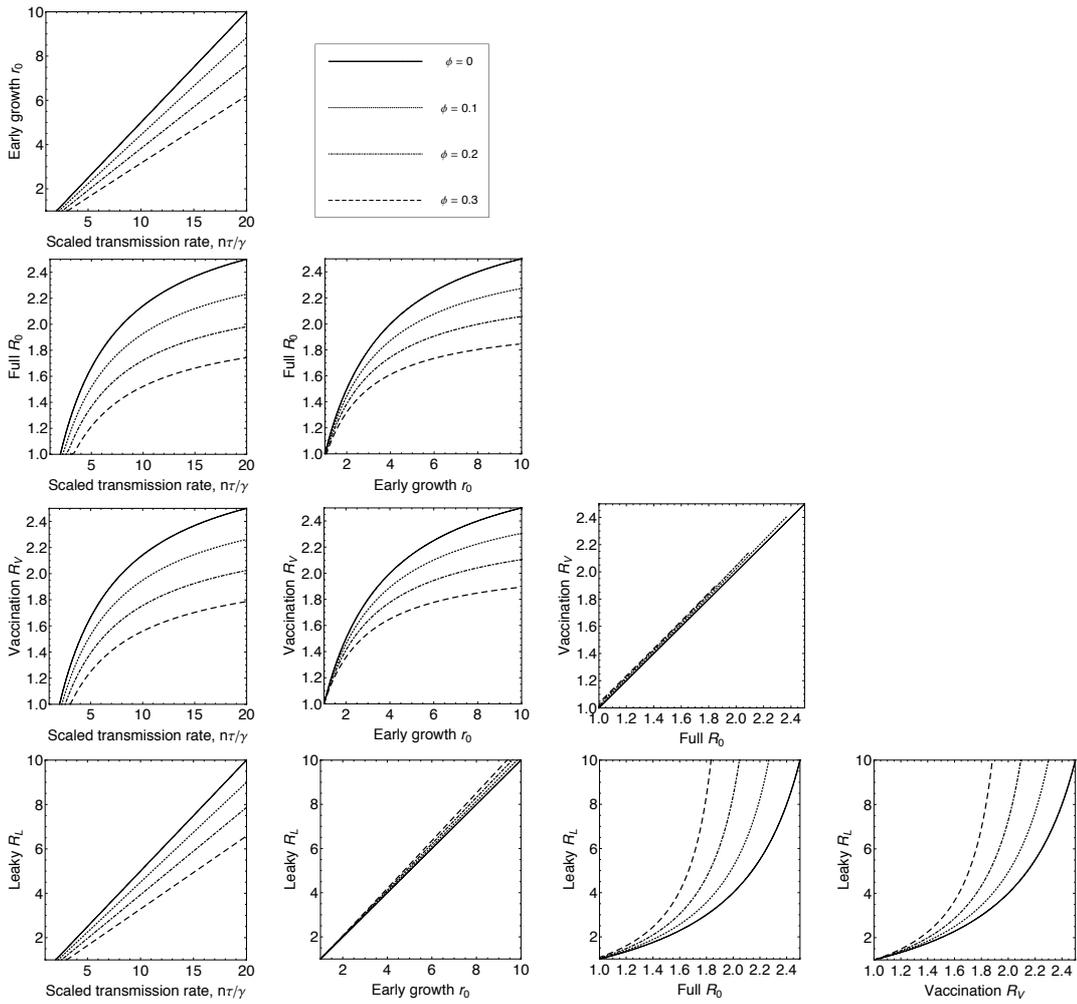} }}
\end{center}
\caption{Observables and reproduction numbers for a clustered network with
$n=4$. The scaled transmission rate $n\tau / \gamma$, early growth $r_0$, full
$R_0$, vaccination $R_V$ and leaky $R_L$ are all plotted against each other,
showing a diversity of relationships.}
\label{fig:rep}
\end{figure}

We now consider numerical integration of the pairwise system~\eqref{pw}. This
requires parameters to be chosen, and so we consider a network with small
neighbourhood size $n=4$ and vary clustering from $0$ to $0.3$.  We consider
two sets of results: in Figure~\ref{fig:rep}, we consider the relationship
between different observable and intervention reproduction numbers; and in
Figure~\ref{fig:outcomes}, we consider the relationship between different
observables and epidemic outcomes.

Looking at Figure~\ref{fig:rep}, we see a consistent ordering $(n\tau / \gamma)
> R_L \geq r_0 \geq R_V \geq R_0$ , as would be expected from exact results for
household models~\cite{Goldstein:2009}. We also find, however, that even when
they do not agree, the early growth observable $r_0$ is strongly predictive of
the leaky vaccination threshold $R_L$, while the basic reproductive ratio $R_0$
is strongly predictive of the random, effective vaccination threshold $R_V$.
This is the case despite the fact that other pairs of reproduction numbers
(e.g.\ $r_0$ and $R_0$) can differ very significantly at different levels of
clustering.

\subsection*{Outcomes}

We now consider the predictive power of different quantities that are
observable early in the epidemic: the transmission rate, early growth rate and
basic reproduction number. Our selection of outcomes is the final size (also
called \textit{attack rate}), which determines the overall proportion of the
population that has suffered from the disease, the peak prevalence, which is
important for assessing the maximum burden of clinical disease during an
epidemic, and the time to peak, which determines how much time is available to
prepare for the peak burden.

\begin{figure}[H]
\begin{center}
\scalebox{0.95}{\resizebox{\textwidth}{!}{ \includegraphics{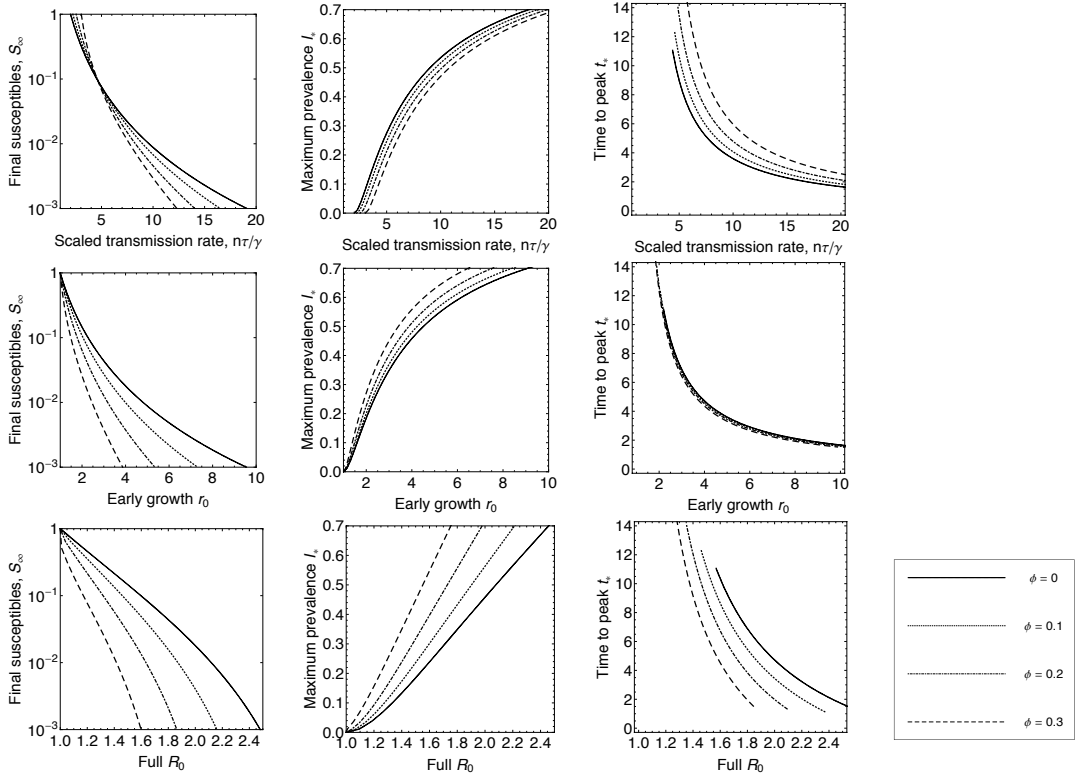} }}
\end{center}
\caption{Outcomes for different observables for a clustered network with $n=4$.
The outcomes of final size, peak height and time are shown in columns, while
rows correspond to constant scaled transmission rate $n\tau / \gamma$, early
growth $r_0$ and full $R_0$.}
\label{fig:outcomes}
\end{figure}

Looking at Figure~\ref{fig:outcomes}, we see several features. At a constant
transmission rate (top row) epidemic peaks are lower and occur later as
clustering is increased. At low transmission rates, clustering decreases attack
rate, but at higher transmission rates, clustering increases attack rate.  This
latter, counter-intuitive effect is seen exact results for
special clustered network types~\cite{House:2010}, but due to the extremely
high attack rates involved, this effect is not easy to reproduce in simulation.
It is possible that this effect may be much larger for different disease
natural histories, and so it may be of practical as well as theoretical
interest.

At constant $r_0$ (middle row) epidemics in clustered populations have larger
attack rates and peaks. While peak times are also slightly reduced, the early
growth rate is strongly predictive of peak time. This predictive power can be
understood by considering the implications of exponential early growth. When an
epidemic peaks, this is due to depletion of susceptibles below the level
required for continued transmission of infection. Exponential growth, by its
nature, places strong bounds on the range of times for which susceptible
depletion can be appreciable and hence the rate of such growth is strongly
predictive of peak time. It is worth noting that the absolute times to peak
will be strongly dependent on the small initial number infectious, which we
take as $I(0) = 10^{-6}$. Where $r_0$ is held constant as $I(0)$ is varied,
different peak time curves will all be shifted right or left by the same
amount, since each epidemic experiences the same early growth rate. For other
rows of Figure~\ref{fig:outcomes}, the shift will need to be determined from
the value of $r_0$ at a given $R_0$ or $n\tau / \gamma$ as shown in
Figure~\ref{fig:rep}.

Similarly to the case of $r_0$, at constant $R_0$ (bottom row) the consequences
of introducing clustering are uniformly larger epidemics, with larger, earlier
peaks. This reversal of outcome prediction for $r_0, R_0$ as compared to
transmission rate can be understood in the following way. Clustering frustrates
the epidemic process early on in an outbreak, as infectious individuals compete
locally for shared susceptibles. Compared to a locally tree-like network,
however, a clustered network offers more routes for infection to travel from
one node to another and so its effect on final outbreak size is non-trivial
to predict.  Once we have adjusted the underlying transmission rate upwards
compared to an unclustered model to give similar $r_0$ or $R_0$, the outbreak
will definitely have a larger size than in the unclustered scenario.

\section*{Discussion}

We have argued in this paper for the merits of a pairwise approach to
understanding epidemics in clustered populations that complements simulations
and exact results for special network structures. Pairwise models allow simply
interpretable conclusions to be drawn with little numerical effort, and involve
a small number of real parameters.

Our results show that for some modelling tasks, clustering does not need to be
considered to get accurate results. In particular, early growth rate predicts
the critical leaky vaccine level and peak time; and $R_0$ predicts the critical
vaccination threshold. For other important calculations, however, the presence
of significant clustering in a population must be modelled to give an accurate
prediction.

The problem of epidemic prediction and model misspecification is, of course,
also posed for many other extensions of the standard \SIR{} model. What appears
to be unique about clustering is that it predicts smaller epidemics at constant
transmission rate and larger epidemics at constant early growth rate as
clustering is increased, in contrast to other forms of population
structure~\cite{Diekmann:2000}, where heterogeneity leads to larger epidemics
at constant transmission rate and smaller epidemics at constant $R_0$.  Another
way of looking at our results is, therefore, that if an epidemic has an attack
rate lower than would be expected from the \SIR{} model, and the population is
clustered, then standard forms of heterogeneity must be even more significant
than an unclustered model would predict---although a mathematical model of the
interaction between clustering and heterogeneity (partially addressed
by~\cite{Volz:2010,House:2010a}) would help to make this intuition clearer.

\section*{Acknowledgements}

Work funded by the UK Medical Research Council (grant number G0701256) and the
UK Engineering and Physical Sciences Research Council (grant number
EP/H016139/1). We would like to thank Istv\'{a}n Kiss and Michael Taylor for
helpful comments on this work.

\newpage

\newpage

\newpage

\end{document}